\documentclass[%
 reprint,
nofootinbib,
nobibnotes,
bibnotes,
amsmath,amssymb,
]{revtex4-2}

\usepackage{graphicx}
\usepackage{dcolumn}
\usepackage{bm}

\begin{document}


\title{
Proof of Spin-Statistics Theorem in Quantum Mechanics 
\\
of Identical Particles
}

\author{Takafumi Kita}
\affiliation{Department of Physics, Hokkaido University, Sapporo 060-0810, Japan}

\date{\today}

\begin{abstract}
A nonrelativistic proof of the spin-statistics theorem 
is given in terms of the field operators satisfying commutation and anticommutation relations, which are introduced here 
in the coordinate space as a means to build the permutation symmetry  into the brackets of identical particles.
An eigenvalue problem of a $\pi$-rotation for a product of two annihilation operators is combined with an analysis on its rotational property to
prove the connection that the field operators
for integral-spin and half-integral-spin particles obey the commutation and anticommutation relations, respectively.
\end{abstract}

\maketitle

\section{Introduction}

Systems of identical particles are characterized by the indistinguishability 
that any virtual exchange of particles does not affect the physical state at all.
This permutation symmetry leads to the profound conclusion 
that the corresponding wave function in quantum mechanics 
should be either symmetric or antisymmetric,
as already stated explicitly in 1926 by 
Heisenberg \cite{Heisenberg26} and Dirac \cite{Dirac26} 
with its relationship to 
Bose-Einstein statistics \cite{Bose24,Einstein24} and the Pauli exclusion principle \cite{Pauli25}.
The statistical-mechanical distribution of identical particles obeying the Pauli exclusion principle
is now called Fermi-Dirac statistics based on their pioneering applications to ideal gases \cite{Fermi26,Dirac26}.

The spin-statistics theorem states the connection of the two kinds of permutation symmetry or statistics with the spin of the constituent particles:
the wave function of identical particles is either symmetric or antisymmetric
according to whether the spin is integral or half-integral.
However, its proofs have been 
carried out mostly in terms of relativistic quantum field theory
with the Lorentz invariance \cite{Fierz39,Belinfante39,PB40,deWet40,Pauli40,Feynman49,Schwinger51,LZ58,Burgoyne58}, notably by Pauli \cite{Pauli40},
under several assumptions such as no negative energies, positive-definite metric, and 
two-kinds of commutation relations at space-like separations \cite{Burgoyne58};
see Refs.\ \onlinecite{DS97} and \onlinecite{CGH12} for a review, historical survey of the topic, and references.
Thus, a simple nonrelativistic proof still remains to be achieved,
just as remarked and regretted by Feynman back in 1963 \cite{Feynman}.
It should be noted in this context that quantum field theory generally starts from introducing 
commutation or anticommutation relations to the fields \cite{Dirac27,Jordan27,JW28}, i.e., the procedure often called {\it second quantization}, 
but its full connection to the permutation symmetry of many-particle wave functions 
seems not to have been recognized to an enough extent.

Thus, we here present a distinct proof in terms of the permutation symmetry.
Unlike previous nonrelativistic attempts to carry it out based on two-particle wave functions \cite{Bacry95,Berry97,Jabs10},
we focus on field operators introduced mathematically 
to realize the permutation symmetry in systems of identical particles.
Combining permutation and rotational symmetries of a product of two annihilation operators, we discuss directly which of the commutation and anticommutation relations 
should be adopted for integral-spin and half-integral-spin particles.

\section{Permutation Symmetry and Field Operators:}

Let us recapitulate the field operators
in quantum mechanics of many-particle systems
introduced solely to describe their permutation symmetry \cite{Kita15},
on which the present proof is based.
We consider a system of $N$ identical particles with mass $m$ and spin $s$.
The Hamiltonian $\hat{H}$ of such a system is characterized by
the symmetry that it is invariant under any permutation
\begin{align}
\hat{P} = \left(
\begin{array}{ccccc}
1 & 2 & 3 & \cdots & N
\\
p_1 & p_2 & p_3 & \cdots & p_N
\end{array}
\right),
\label{hatP}
\end{align}
where $1\leq p_j\leq N$ with no duplication among the $p_j$'s. 
This may be seen by considering a standard Hamiltonian
\begin{align}
\hat{H}=\sum_{j=1}^N \left[\frac{\hat{\bf p}_j^2}{2m}+U({\bf r}_j)\right]+\sum_{j=1}^{N-1} \sum_{j'=j+1}^N V(|{\bf r}_j-{\bf r}_{j'}|),
\label{H}
\end{align}
where $\hat{\bf p}_j$ and ${\bf r}_j$ are the momentum operator and position, respectively, and $U({\bf r}_j)$ and $V(|{\bf r}_j-{\bf r}_{j'}|)$ denote  one-particle and interaction potentials, respectively.
Operation $\hat{P}\hat{H}\hat{P}^{-1}$ on Eq.\ (\ref{H}) merely changes the order of the summation over $j$
and $(j,j')$, thus satisfying $\hat{P}\hat{H}\hat{P}^{-1}=\hat{H}$. Writing it as
$\hat{P}\hat{H}=\hat{H}\hat{P}$,
we realize that: (i) $\hat{P}$ and $\hat{H}$
can be diagonalized simultaneously,
and (ii) any expectation value of $\hat{P}$ does not change in time according to the Heisenberg equation of motion \cite{Sakurai}.
Hence, the eigenvalue problems of $\hat{H}$ and $\hat{P}$ 
for the $N$-particle wave functions,
\begin{align}
\hat{H}\Phi_\nu(\xi_1,\xi_2,\cdots,\xi_N)=&\,{\cal E}_\nu\Phi_\nu(\xi_1,\xi_2,\cdots,\xi_N),
\label{H-eigen}
\\
\hat{P}\Phi_\nu(\xi_1,\xi_2,\cdots,\xi_N)=&\,\sigma^P \Phi_\nu(\xi_1,\xi_2,\cdots,\xi_N)
\label{P-eigen}
\end{align}
can be solved simultaneously, where $\xi_j\equiv ({\bf r}_j,m_{{\rm s}j})$ collectively 
specifies the full quantum-mechanical coordinates with $m_{{\rm s}j}$ denoting 
an eigenvalue of the spin operator $\hat{s}_{zj}$.
The eigenvalue $\sigma^P$ of Eq.\ (\ref{P-eigen}) can be written in terms of 
\begin{align}
\sigma =\pm 1 
\label{sigma}
\end{align}
as
\begin{align}
\sigma^P = \left\{\begin{array}{ll} 
1 & \mbox{when $\hat{P}$ is even} \\
\sigma & \mbox{when $\hat{P}$ is odd}
\end{array}\right. .
\label{sigma^P}
\end{align}
The eigenfunction of Eq.\ (\ref{P-eigen}) with Eq.\ (\ref{sigma^P}) is  
either completely symmetric ($\sigma=1$) or antisymmetric ($\sigma=-1$).
Excluding any other possibility for wave functions of identical particles is sometimes called {\it symmetrization postulate} \cite{MG64}, which is fully accepted here from the beginning
based on the fact that no manifest experiment that contradicts it 
has been reported.
Our purpose here is to connect the sign of $\sigma$ with the magnitude of $s$.

The eigenvalue problem of Eq.\ (\ref{P-eigen}) for permutations can be solved as follows.
First, we define the commutator
\begin{align}
[\hat{A},\hat{B}]_\sigma\equiv \hat{A}\hat{B}-\sigma\hat{B}\hat{A}
\label{[A,G]}
\end{align}
of general operators $\hat{A}$ and $\hat{B}$ with the $\sigma$ dependence.
It should be noted that the subscript $\sigma$ in Eq.\ (\ref{[A,G]}) is opposite 
in sign to the convention used in relativistic quantum field theory \cite{Dirac27,Pauli40,Schwinger51,DS97}.
However, the present notation has the advantage 
in many-particle physics that it connects the sign of $\sigma$ in the permutation directly to the commutator.
Using Eq.\ (\ref{[A,G]}), we introduce the field operators $\hat\psi(\xi)$ and $\hat\psi^\dagger(\xi)$
that satisfy the commutation relations
\begin{subequations}
\label{psi-commute} 
\begin{align}
\bigl[\hat{\psi}(\xi) ,
\hat{\psi}^{\dagger}(\xi') \bigr]_\sigma
=&\, \delta(\xi,\xi') ,
\label{psi-commute1} 
\\
\bigl[\hat{\psi}(\xi) ,\hat{\psi}(\xi') \bigr]_\sigma =&\,
\bigl[\hat{\psi}^{\dagger}(\xi) ,
\hat{\psi}^{\dagger}(\xi') \bigr]_\sigma= 0 
\label{psi-commute2}
\end{align}
\end{subequations}
with $\delta(\xi,\xi') \equiv\delta({\bf r}-{\bf r}')\delta_{m_{{\rm s}}m_{{\rm s}}'}$.
In addition, we define the ket $|0\rangle$ and bra $\langle 0 |$ via the right (left) action of the annihilation (creation) operator,
\begin{equation}
\hat{\psi}(\xi) |0\rangle = 0 , \hspace{3mm}
0=\langle 0 | \hat{\psi}^{\dagger}(\xi)= \bigl[\hat{\psi}(\xi) |0\rangle\bigr]^* ,
 \hspace{3mm}
\langle 0 | 0\rangle = 1 .
\label{|0>}
\end{equation}
These are the basic ingredients needed to construct the eigenspace of permutations
mathematically.

Using the field operators and brackets, we can create the eigenstates by
\begin{subequations}
\label{|xi_1xi_2...xi_N>_tot}
\begin{align}
|\xi_1,\xi_2,\cdots,\xi_N\rangle \equiv \frac{1}{\sqrt{N!}}\hat\psi^\dagger(\xi_1)\hat\psi^\dagger(\xi_2)
\cdots \hat\psi^\dagger(\xi_N)|0\rangle .
\label{|xi_1xi_2...xi_N>}
\end{align}
For example, $|\xi_2,\xi_1,\xi_3,\cdots,\xi_N\rangle=\sigma|\xi_1,\xi_2,\xi_3,\cdots,\xi_N\rangle$ holds, as can be shown easily by using Eq.\ (\ref{psi-commute2}),
and it generally satisfies $\hat{P}|\xi_1,\xi_2,\cdots,\xi_N\rangle=\sigma^P
|\xi_1,\xi_2,\cdots,\xi_N\rangle$.
The conjugate $\langle \xi_1,\xi_2,\cdots,\xi_N|\equiv |\xi_1,\xi_2,\cdots,\xi_N\rangle^*$ of Eq.\ (\ref{|xi_1xi_2...xi_N>}) is given by
\begin{align}
\langle \xi_1,\xi_2,\cdots,\xi_N|=&\, \frac{1}{\sqrt{N!}}\langle 0| \hat\psi(\xi_N)\cdots \hat\psi(\xi_2)
\hat\psi(\xi_1).
\label{|xi_1xi_2...xi_N>^*}
\end{align}
\end{subequations}
Using Eqs.\ (\ref{psi-commute}) and (\ref{|0>}), one can show that the 
brackets of Eq.\ (\ref{|xi_1xi_2...xi_N>_tot}) fulfill the orthonormality
\begin{subequations}
\begin{align}
&\,\langle \xi_{1}',\xi_{2}',\cdots,\xi_{N'}'|\xi_{1},\xi_{2},\cdots,\xi_{N}\rangle
\notag \\
=&\,\frac{\delta_{N'N}}{N!}\sum_{\hat{P}}\sigma^{P}\delta(\xi_{1}',\xi_{p_{1}})
\delta(\xi_{2}',\xi_{p_{2}})\cdots \delta(\xi_{N}',\xi_{p_{N}}).
\label{ortho}
\end{align}
They also satisfy the completeness relation
\begin{align}
\int \! d\xi_{1} \int \! d\xi_{2} \cdots\int  \! d\xi_{N}\,
|\xi_{1},\xi_{2},\cdots,\xi_{N}\rangle
\langle \xi_{1},\xi_{2},\cdots,\xi_{N}| =1 
\label{complete}
\end{align}
\end{subequations}
within the eigenspace of permutations, where the integration over $\xi_j$ denotes
\begin{align}
\int \! d \xi_j \equiv \sum_{m_{{\rm s}j}=-s}^s \int\! d^3 r_j .
\label{IntDxi}
\end{align}
More generally, the left-hand side of Eq.\ (\ref{complete}) forms the projection operator
onto the eigenspace of permutations.
Thus, the eigenvalue problem (\ref{P-eigen}) has been solved completely 
based on Eqs.\ (\ref{psi-commute}) and (\ref{|0>}).

The simultaneous eigenfunction of Eqs.\ (\ref{H-eigen}) and (\ref{P-eigen}) can be written 
in terms of Eq.\ (\ref{|xi_1xi_2...xi_N>^*}) as
\begin{align}
\Phi_\nu(\xi_1,\xi_2,\cdots,\xi_N)=\langle \xi_1,\xi_2,\cdots,\xi_N|\Phi_\nu\rangle .
\label{Psi}
\end{align}
Thus, the wave function denotes the full coordinate representation of 
the many-particle ket $|\Phi_\nu\rangle$. 
Moreover, the Schr\"odinger equation (\ref{H-eigen}) can be written concisely 
by using $|\Phi_\nu\rangle$ as
\begin{align}
\hat{\cal H}|\Phi_\nu\rangle={\cal E}_\nu|\Phi_\nu\rangle,
\label{Sch-ket}
\end{align}
where $\hat{\cal H}$ for the specific example of Eq.\ (\ref{H}) is given by
\begin{align}
\hat{\cal H}=&\,\int d\xi\, \hat\psi^\dagger(\xi)\left[\frac{\hat{\bf p}^2}{2m}+U({\bf r})\right]\hat\psi(\xi)
\notag \\
&\,
+\frac{1}{2}\int d\xi \int d\xi' \, V(|{\bf r}-{\bf r}'|)
\hat\psi^\dagger(\xi)\hat\psi^\dagger(\xi')
\hat\psi(\xi')\hat\psi(\xi) .
\label{calH}
\end{align}
One can show that operating $\langle \xi_1,\xi_2,\cdots,\xi_N|$ from the left of Eq.\ 
(\ref{Sch-ket}) reproduces Eq.\ (\ref{H-eigen}).
Also integrating Eq.\ (\ref{Psi}) over $\xi_j$ for $j=n+1,\cdots,N$, 
we obtain an $n$-body correlation function of the ket $|\Phi_\nu\rangle$, which for $n=2$ is given explicitly by
\begin{align}
F_\nu(\xi_1,\xi_2)\equiv \int d\xi_3\cdots\int d\xi_N \,
\langle \xi_1,\xi_2,\xi_3,\cdots,\xi_N|\Phi_\nu\rangle .
\label{F-def}
\end{align}
This function plays a key role in our proof.
The squared amplitude  $|F_\nu(\xi_1,\xi_2)|^2$ denotes what is called the {\it pair distribution function}
in the literature.

To see the connection of the two kinds of permutation symmetry with
Bose-Einstein and Fermi-Dirac statistics, 
let us focus on ideal gases with $V=0$ in Eq.\ (\ref{calH}).
We then expand the field operators 
in terms of the eigenstates $\{\phi_q(\xi)\}_q$ of the one-particle Schro\"dinger equation
\begin{align}
\left[\frac{\hat{\bf p}^2}{2m}+U({\bf r})\right] \phi_q(\xi)=\varepsilon_q \phi_q(\xi),
\end{align}
as
\begin{align}
\hat{\psi}(\xi)=\sum_q \hat{c}_q\phi_q(\xi) .
\end{align}
One can show based on Eq.\ (\ref{psi-commute}) that $\hat{c}_q$ satisfies
\begin{align}
[\hat{c}_q,\hat{c}_{q'}^\dagger]_\sigma=\delta_{qq'},\hspace{5mm}
[\hat{c}_q,\hat{c}_{q'}]_\sigma=[\hat{c}_q^\dagger,\hat{c}_{q'}^\dagger]_\sigma=0 .
\label{c_k-commute}
\end{align}
The Hamiltonian (\ref{calH}) with $V=0$, which we denote by $\hat{\cal H}_0$, can be written as
\begin{align}
\hat{\cal H}_0=\sum_q \varepsilon_q\hat{c}_q^\dagger\hat{c}_q .
\label{H_0-diag}
\end{align}
Every eigenstate $\nu$ of Eq.\ (\ref{H_0-diag}) is specified completely in terms of the number $n_{q}$ of particles in each one-particle state $q$ as
\begin{align}
|\Phi_{\nu}\rangle\equiv &\,|n_{q_1},n_{q_2},n_{q_3},\cdots\rangle
\notag \\
\equiv &\,\frac{\left(\hat{c}_{q_{1}}^{\dagger}\right)^{n_{q_1}}}{\sqrt{n_{q_1}!}}\frac{\left(\hat{c}_{q_2}^{\dagger}\right)^{n_{q_2}}}{\sqrt{n_{q_2}!}}
\frac{\left(\hat{c}_{q_3}^{\dagger}\right)^{n_{q_3}}}{\sqrt{n_{q_3}!}}\cdots\cdots
|0\rangle .
\label{|Psi-nonint>}
\end{align}
It follows from Eq.\ (\ref{c_k-commute}) that possible values for $n_{q}$ differ between $\sigma=\pm 1$ as
\begin{equation}
n_{q}=\left\{
\begin{array}{ll}
0,1,2,\cdots & \,\,\,\,(\sigma =+1)
\\
0,1 & \,\,\,\,(\sigma =-1)
\end{array}
\right. ,
\label{n_k-0}
\end{equation}
which represents Bose-Einstein $(\sigma =+1)$ and Fermi-Dirac
$(\sigma =-1)$ statistics.
Finally, particle number $N$ and energy ${\cal E}_\nu$ of state $\nu$ are expressible as
\begin{equation}
N=\sum_{q} n_{q} ,\hspace{5mm}
{\cal E}_\nu = \sum_{q} n_{q} \varepsilon_{q} .
\label{NE-ideal}
\end{equation}
Thus, the field operators obeying the commutation relations
(\ref{psi-commute2}) are quite useful for describing systems of identical particles
with permutation symmetry. 
The occupation-number representation of Eq.\ (\ref{|Psi-nonint>}) without interaction is also 
called {\it Fock states} \cite{Fock32}, from which the second quantization in many-particle physics generally starts \cite{AGD63,FW71,NO88,Mahan00} similarly as in relativistic quantum field theory \cite{Dirac27,Jordan27,JW28}.
However, the present construction based on Eqs.\ (\ref{psi-commute}) and (\ref{|0>}),
which focuses on the permutation symmetry, is advantageous in that it is manifestly applicable to systems with interactions such as superconductivity with phase coherence \cite{Kita15}.

\section{Rotational Symmetry of Field Operators:}

We now focus on the rotational symmetry of the field operator $\hat{\psi}(\xi)$,
which depends crucially on spin $s$.
To this end, it is convenient to separate the spin argument $m_{{\rm s}}$ from
$\xi=({\bf r},m_{{\rm s}})$ as $\hat{\psi}(\xi)=\hat\psi_{m_{{\rm s}}}({\bf r})$
and express $m_{{\rm s}}=s,s-1,\cdots, -s$ components as a column vector
\begin{align}
\hat{\bm\psi}({\bf r})\equiv\begin{bmatrix}
\hat\psi_s({\bf r}) \\ \hat\psi_{s-1}({\bf r}) \\ \vdots \\ \hat\psi_{-s}({\bf r})
\end{bmatrix}.
\label{bmPsi}
\end{align}
Upon rotating the coordinate system by $\theta$ about the $z$ axis,
Eq.\ (\ref{bmPsi}) is transformed into $\hat{R}_\theta\hspace{0.3mm}\hat{\bm\psi}({\bf r})$ with
\begin{align}
\hat{R}_\theta\hspace{0.3mm}\hat{\bm\psi}({\bf r})\equiv \exp\bigl(i\theta \underline{s}_z\bigr)\hspace{0.3mm}\hat{\bm\psi}(\underline{R}^{-1}_\theta\,{\bf r})
\label{hatRpsi}
\end{align}
where $\underline{s}_z$ is the $(2s+1)\times (2s+1)$ matrix whose elements are given by
$\langle m_{{\rm s}} |\hat{s}_z|m_{{\rm s}}'\rangle =m_{{\rm s}}\delta_{m_{{\rm s}}m_{{\rm s}}'}$, and $\underline{R}_\theta^{-1}$
denotes the inverse of the matrix
\begin{align}
\underline{R}_\theta\equiv \begin{bmatrix} \cos\theta & \sin\theta & 0 \\
-\sin\theta & \cos\theta & 0\\
0 & 0 & 1
\end{bmatrix}.
\end{align}
A rotation about a tilted axis can be defined similarly
in terms of the Euler angles \cite{ITO90},
but Eq.\ (\ref{hatRpsi}) suffices for the present purpose.
Equation (\ref{bmPsi}) with Eq.\ (\ref{hatRpsi}) for a half-integral $s$ 
extends the spinor of $s=\frac{1}{2}$ 
discovered by Pauli in 1927 \cite{Pauli27,Tomonaga98} to general half-integers.
The $m_{{\rm s}}$ element of Eq.\ (\ref{hatRpsi}) is given explicitly by
$\hat{R}_\theta\hspace{0.3mm} \psi_{m_{{\rm s}}} ({\bf r})= e^{im_{{\rm s}}\theta}\psi_{m_{{\rm s}}} (\underline{R}_\theta^{-1}{\bf r})$.

\section{Proof of Spin-Statistics Theorem}

We now introduce the operator
\begin{align}
\hat{F}_{m_{{\rm s}},\sigma}({\bf r})\equiv \hat\psi_{m_{{\rm s}}}(-{\bf r})\hat\psi_{m_{{\rm s}}}({\bf r}) ,
\label{calO}
\end{align}
which is relevant to two-body correlations of equal-spin particles.
Indeed, it follows from Eq.\ (\ref{F-def}) with $(\xi_1;\xi_2)=({\bf r},m_{\rm s};-{\bf r},m_{\rm s})$
that we can clarify the relative motion of two particles in terms of Eq.\ (\ref{calO}). 
Our proof of the spin-statistics theorem 
proceeds with the above product of two annihilation operators,
in contrast to the relativistic ones performed in terms of creation-annihilation pairs \cite{Fierz39,Belinfante39,PB40,deWet40,Pauli40,Feynman49,Schwinger51,LZ58,Burgoyne58,DS97,CGH12}.
It should be noted that the origin ${\bf r}={\bf 0}$ and $z$ axis in the coordinate system of Eq.\ (\ref{calO})
can be any position and direction in space, respectively, as realized by the translation and rotation
of the coordinate system.

Let us clarify the symmetries of Eq.\ (\ref{calO}). 
First, it satisfies
\begin{subequations}
\label{calO-symm}
\begin{align}
\hat{F}_{m_{{\rm s}},\sigma}(-{\bf r})=\sigma \hat{F}_{m_{{\rm s}},\sigma}({\bf r})
\label{calO-symm1}
\end{align}
because of Eq.\ (\ref{psi-commute2}).
Second, operating $\hat{R}_\theta$ on Eq.\ (\ref{calO}) 
at ${\bf r}_\perp\equiv (x,y,0)$ yields
\begin{align}
\hat{R}_\theta \hat{F}_{m_{{\rm s}},\sigma}({\bf r}_\perp)=e^{2im_{{\rm s}}\theta} \hat{F}_{m_{{\rm s}},\sigma}(\underline{R}_\theta^{-1}{\bf r}_\perp) 
\label{calO-symm2}
\end{align}
\end{subequations}
as seen from Eq.\ (\ref{hatRpsi}). 
These are the basic symmetries of $\hat{F}_{m_{{\rm s}},\sigma}({\bf r})$.

Setting $\theta=\pi$ in Eq.\ (\ref{calO-symm2}) yields
\begin{align*}
\hat{R}_\pi\, \hat{F}_{m_{{\rm s}},\sigma}({\bf r}_\perp) =
(-1)^{2s} \,\hat{F}_{m_{{\rm s}},\sigma}(-{\bf r}_\perp),
\end{align*}
which can be transformed by using Eq.\ (\ref{calO-symm1}) into the eigenvalue problem
\begin{align}
\hat{R}_\pi\, \hat{F}_{m_{{\rm s}},\sigma}({\bf r}_\perp)
=\lambda \,\hat{F}_{m_{{\rm s}},\sigma}({\bf r}_\perp),
\label{EigenR_pi}
\end{align}
with 
\begin{align}
\lambda\equiv (-1)^{2s} \sigma .
\label{lambda-def}
\end{align}
It follows from $\hat{R}_\pi^2=1$ that $\lambda^2=1$ holds so that $\lambda$
can be either $+1$ or $-1$.
Thus, Eq.\ (\ref{EigenR_pi}) with Eq.\ (\ref{lambda-def}) 
shows a manifest spin-statistics connection in quantum mechanics
of identical particles.
Though $\lambda=\pm 1$ may be regarded as trivial from $|(-1)^{2s}|=|\sigma|=1$,
Eq.\ (\ref{lambda-def}) has the manifest advantage of specifying $\lambda=(-1)^{2s} \sigma$ as an eigenvalue of the $\pi$-rotation.

First, let us focus on the case of a half-integral $s$, where Eq.\ (\ref{EigenR_pi}) reduces to 
$\hat{R}_\pi \hat{F}_{m_{{\rm s}},\sigma}({\bf r}_\perp) =-\sigma \hat{F}_{m_{{\rm s}},\sigma}({\bf r}_\perp)$.
Moreover, Eq.\ (\ref{calO-symm2}) tells us that 
${\bf r}_\perp={\bf 0}$ is a singular point around which 
$\hat{F}_{m_{{\rm s}},\sigma}({\bf r}_\perp)$ changes its phase by $4\pi m_{\rm s}$ ($=\pm 2\pi,\pm 4\pi,\cdots$) upon a counterclockwise rotation about the $z$ axis.
Hence, $\hat{F}_{m_{{\rm s}},\sigma}({\bf 0})$ should vanish identically to meet the requirement that the wave function be single-valued.
However, it is not automatically satisfied for $\sigma=1$ where $\hat{F}_{m_{{\rm s}},+}({\bf 0})$ can be finite according to Eqs.\ (\ref{calO}) and (\ref{calO-symm1}). 
Thus, we conclude that the equality
\begin{align}
(-1)^{2s} \sigma=1
\label{Spin-Statistics-Theorem}
\end{align}
should hold,
i.e., $\sigma=-1$ for any half-integral $s$.

Next, we consider the case of an integral $s$, where 
Eq.\ (\ref{EigenR_pi}) is simplified to
$\hat{R}_\pi \hat{F}_{m_{{\rm s}},\sigma}({\bf r}_\perp) =\sigma \hat{F}_{m_{{\rm s}},\sigma}({\bf r}_\perp)$.
This is the usual relation to distinguish between even ($\sigma=1$) and odd ($\sigma=-1$)
functions.
However, the latter choice $\sigma=-1$ brings the severe restriction that no wave function even in ${\bf r}$
is possible for any relative motion of two particles,
even in the presence of an attractive interaction between them.
With no obvious or rational reason for the vanishing of any relative wave function at ${\bf r}={\bf 0}$, we should take $\sigma=1$ for integral-spin particles, which is also supported experimentally.
Adopting it, we arrive at Eq.\ (\ref{Spin-Statistics-Theorem}) once again.
In conclusion, Eq.\ (\ref{Spin-Statistics-Theorem}) holds true irrespective
of spin. This completes our proof.

\section{Cocluding Remarks}

We have presented an elementary proof of  the spin-statistics theorem within quantum mechanics of identical particles based on the field operators
 introduced so as to describe the permutation symmetry of many-particle wave functions.
This approach has enabled us to choose a product of two annihilation operators,
instead of creation-annihilation pairs in the previous relativistic studies,
as the basis of our proof through clarifying two-body correlations and their rotational symmetry.
One may wonder why this kind of proofs couldn't have been performed much earlier.
One of the reasons for it may be traced 
to the general introduction of the field operators
in terms of the non-interacting number representation of Eq.\ (\ref{|Psi-nonint>}) in 
both quantum field theory \cite{Dirac27,Jordan27,JW28} and many-particle physics \cite{AGD63,FW71,NO88,Mahan00},
in which its connection with the permutation symmetry 
may not have been recognized explicitly.
It should be noted finally that the theorem in the context of many-particle physics
is more about the connection between spin and permutation symmetry 
than about spin and statistics.

\end{document}